# Pairing type symmetry in cuprate superconductors

T.B. Charikova[a], N.G.Shelushinina[a], M.R. Popov[a,*]

[a] *Institute of Physics of Metals, Ural Branch, Russian Academy of Sciences, Ekaterinburg, 620108 Russia*
[*] *e-mail: popov_mr@imp.uran.ru*

**Abstract** - We report a brief overview of a number of results on a type of pairing state symmetry in cuprate HTSC from a current literature. The results of recent high-precision measurements of near-nodal *d*-type gap functions by a modern laser-based ARPES method or by a *c*-axis twisted Josephson junction for hole-doped Bi-2212 system are presented. Recent experimental evidence in favor of *d*-type pairing symmetry, with a special nonmonotonic functional form, for electron-doped cuprate NdCeCuO is discussed. A new essential result on the identification of $d_{x^2-y^2}$ symmetry of superconducting order parameter for parent high-$T_c$ cuprate structure, electron-doped infinite layer SrLaCuO system, by the phase sensitive tests is described. A complex of the data presented indicates a vital role of *d*-wave pairing symmetry, in the special $d_{x^2-y^2}$ form, for high-temperature conductivity in cuprate systems.

## 1. INTRODUCTION

Since the discovery of high transition temperature (high-$T_c$) superconductivity in cuprates [1], tremendous work has been done on these materials. It became possible to increase $T_c$ from initially 30 K to 135 K by synthesizing increasingly complex compounds. However, the microscopic mechanism causing high-$T_c$ superconductivity still has not been identified and it is one of the basic issues in solid state physics (see monographs [2–4]).

The determination of the order-parameter symmetry is a critical challenge in identifying the pairing mechanism for high-temperature superconductivity. A growing set of theoretical calculations and experiments have suggested that the high-temperature cuprates may exhibit an unconventional pairing state: a state with an order parameter or energy gap that has a symmetry in momentum space different from that of the isotropic *s*-wave Cooper pair state that is believed to describe almost all low-temperature superconductors.

Much of the recent attention has been concentrated on states with *d*-wave symmetry and specifically, on the $d_{x^2-y^2}$ pairing state [5], [6]. This state has two key properties that make it a possible candidate to explain the behavior of cuprates. Firstly, it reveals nodes in the energy gap, leading to an excess of excitations at low temperatures which corresponds to experimental data. Second, this special *d*-wave symmetry is implied by a number of possible superconducting pairing mechanisms, especially those involving magnetic interactions, which are known to be important in cuprates.

Whereas for hole-doped cuprates $d_{x^2-y^2}$ - wave pairing has been established [5], [6], for electron-doped cuprates the issue has not yet been completely resolved. After controversial discussion, the electron-doped T′-compounds $L_{2-x}Ce_xCuO_4$ (L = La, Pr, Nd, Eu or Sm) have been shown to be predominant *d*-wave superconductors by a number of phase-sensitive experiments [7].

Relying on the results of fundamental reviews [5], [6], [7] and on a number of new data we present a brief survey of some of the key phase-sensitive tests for evidence of *d*–symmetry of superconducting order parameter (SOP) in hole- and electron-doped cuprate HTSC.

The plan for this work is as follows. Recent experimental advances for high-precision determination of SOP symmetry in the hole - doped Bi – system are presented in section 2, new results on definition of SOP symmetry in the infinite layer cuprate structure SrLaCuO are given in section 3, current views on SOP symmetry in electron-doped cuprates PrLaCeCuO and NdCeCuO are set out in section 4 and the contents are summarized in section 5.

## 2. Hole - doped cuprates

The early and recent developments have led to the conclusion that the superconducting gap in the hole-doped cuprates is characterized by a strong momentum dependence and the superconducting gap function is consistent with an order parameter having $d_{x^2-y^2}$ symmetry, with support from ARPES, penetration depth, Raman and phase sensitive measurements [5], [6], [8]. The *d*-wave symmetry of the superconducting gap has become an accepted fact when one constructs theories and interprets experimental results.

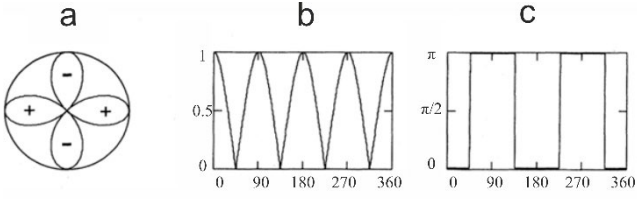

**Fig.1.** Paring state of the $d_{x^2-y^2}$ symmetry (a), magnitude (b) and phase (c) of the superconducting order parameter as a function of direction in CuO$_2$ planes of the cuprates for the $d_{x^2-y^2}$ pairing symmetry (after [5]).

The energy gap (order parameter) of the $\boldsymbol{d_{x^2-y^2}}$ state has the functional form in $\boldsymbol{k}$ space (see, e.g., [5]):

$$\Delta(\boldsymbol{k}) = \Delta_0[\cos(k_x a) - \cos(k_y a)], \qquad (1)$$

where $\Delta_0$ is the maximum gap value and $a$ is the in-plane lattice constant.

The gap is real with a strongly anisotropic magnitude featuring nodes along the (110) directions in $k$ space and a sign change in the order parameter between the sections in the $k_x$ and $k_y$ directions. Physically, this sign change indicates a relative phase of $\pi$ in the superconducting condensate wave function for Cooper pairs with orthogonal relative momenta. In the cuprates, this state is believed to describe the order parameter in the CuO$_2$ planes, with the sections being aligned with the in-plane lattice vectors (see Fig. 1).

In the hole-doped high-$T_c$ cuprate superconductors a single band, originating mainly from hybridized Cu $3d_{x^2-y^2}$ and O $2p_{x,y}$ orbitals, crosses the Fermi level ($E_F$) forming a large hole-like Fermi surface. On the Fermi surface (Fig. 2a), the gap is the largest at the antinode Fermi momentum ($k_F$) on the Brillouin zone boundary near $(\pi, 0)$, where the Fermi angle $\theta = 0^0$. The gap size gradually decreases towards the node along the Fermi surface and becomes zero at the node $k_F$ in the CuO bond diagonal direction ($\theta = 45^0$). The superconducting gap changes sign across the node (Fig. 2b).

### 2.1. Angle-resolved photoemission spectroscopy (ARPES)

The momentum-resolved nature of ARPES makes it a leading powerful tool in study the spectral gaps and establishing the anisotropic $d$-wave structure of the superconducting gap for high-transition-temperature cuprates in contrast with the conventional isotropic $s$-wave superconducting gap. The article of Hashimoto et al. [8] overviews the current ARPES results with significantly improved instrumentation and carefully matched experiments in recent years (see Fig. 3a,b).

The cuprates are well-suited for the ARPES technique because of their quasi-2D electronic structure. Over the past two decades, experiments have improved tremendously, allowing more precise information about electronic structure, including the gap functions, to be obtained, in particular, for Bi$_2$Sr$_2$CaCu$_2$O$_{8+\delta}$ (Bi2212) and Bi$_2$Sr$_{2-x}$La$_x$CuO$_{6+\delta}$ (Bi2201) families.

One recent development is the use of narrow-bandwidth ultraviolet lasers as light sources for photoemission [9]. The superior resolution of laser ARPES provides unprecedented access to the lowest energy excitations near the node, as shown in Fig. 3 a,b.

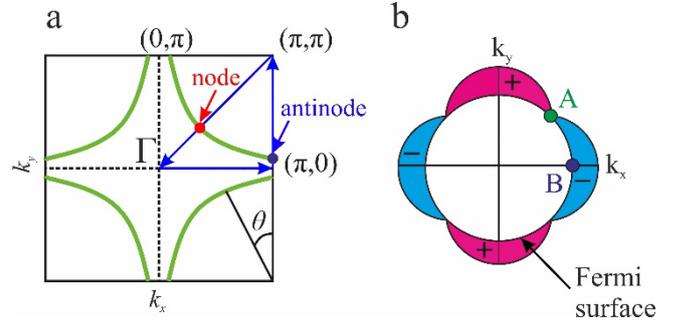

**Fig. 2.** *d*-wave superconducting gap symmetry in cuprates. (a) Hole-like Fermi surface in cuprate superconductor, the nodal and antinodal momenta and the Fermi angle θ are defined; (b) Schematic of a *d*-wave order parameter on a circular Fermi surface. Gap is zero at the node where the superconducting gap changes sign (A) and maximum at the antinode (B) (after [8]).

Furthermore, traditional synchrotron-based ARPES continues to be improved with brighter synchrotrons and more powerful spectrometers. When one combines modern synchrotron and laser-based ARPES experiments, one can gain deep insights into the nature of energy gaps, as reviewed in [8].

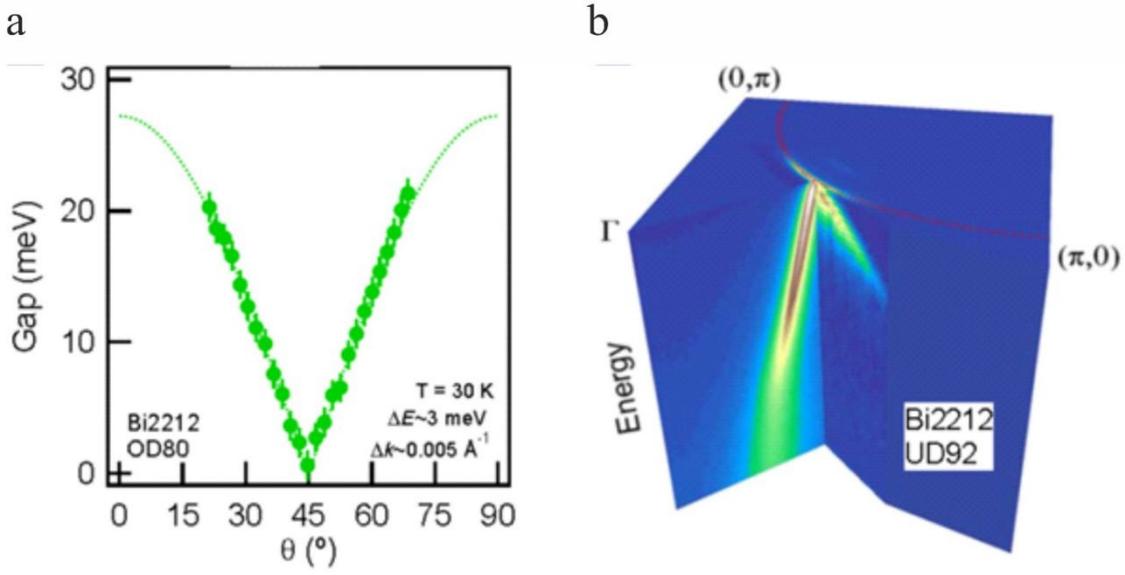

**Fig. 3.** *d*-wave superconducting gap symmetry in Bi2212 witnessed by ARPES. (a) Near-nodal gaps measured by a modern laser-based ARPES system with superior resolutions. (b) Three dimensional ARPES data set, showing the quasiparticle dispersions both perpendicular and parallel to the Fermi surface near the node, reproduced from [9] (after [8]).

2.2. Twisted Josephson junctions ("twistronics")

Weak van der Waals (vdW) bonding between neighboring atomic layers offers a unique opportunity for engineering atomic interfaces with controlled twist angles. Within the confines of such *twistronics* electronic states at the twisted interface of various vdW materials, including graphene [10], can be investigated.

coupling between nodal superconducting order parameters across a twisted vdW interface (see, e.g., [11]). Twisted interfacial Josephson junctions between superconductors directly probe the pairing symmetry of Cooper pairs.

The preservation of surface superconductivity of BSCCO crystals after vdW stacking remains an outstanding experimental challenge. Employing a novel cryogenic assembly technique, Josephson junctions with an atomically sharp twisted interface

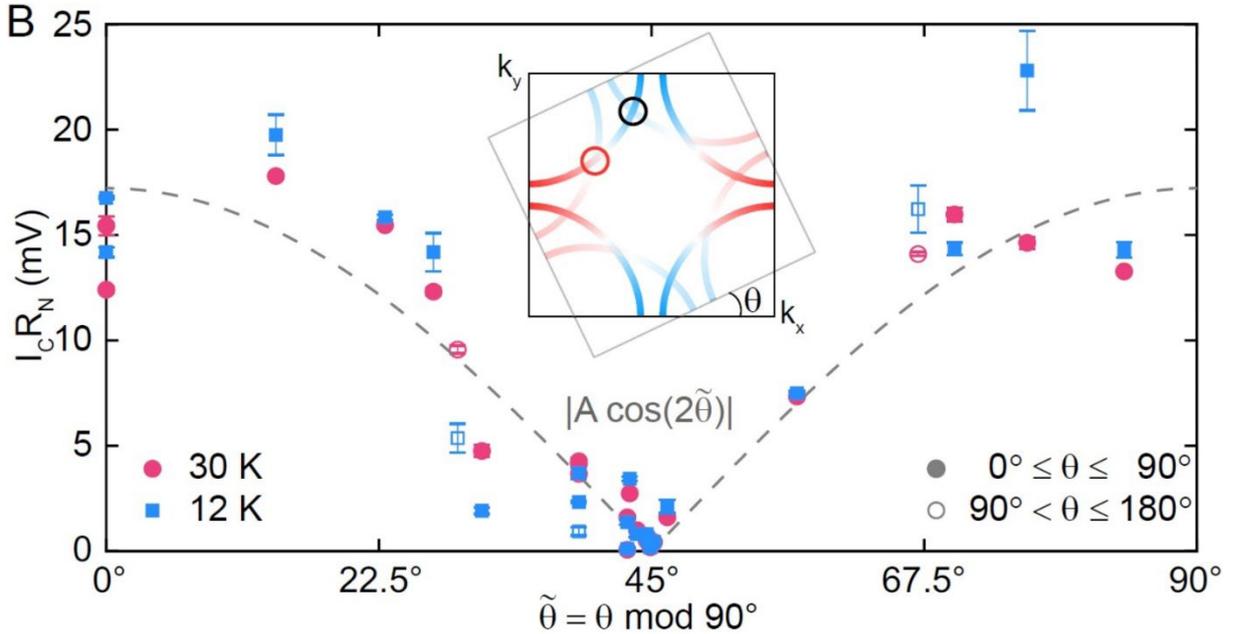

**Fig. 4.** *d*-wave symmetry of superconducting order parameter revealed by supercurrent tunneling. Angular dependence of $I_C R_N$ for 24 devices at 30 and 12 K. The points follow the $|\cos(2\tilde{\theta})|$ curve predicted for nearly incoherent tunneling between *d*-wave superconductors (after [12]).

Layered cuprate high temperature superconductors also offer a platform for twistronics by engineering the

between $Bi_2Sr_2CaCu_2O_{8+x}$ (BSCCO) crystals were fabricated [12]. In BSCCO, superconducting $CuO_2$

bilayers are Josephson-coupled through insulating [SrO-BiO] bilayers, where the crystal can be mechanically cleaved into atomically flat crystals exhibiting superconductivity.

The interfacial Josephson coupling between twisted nodal d-wave superconductors is strongly modulated by the twist angle [11]. The Josephson critical current density $I_c$ sensitively depends on the twist angle, reaching the maximum value comparable to that of the intrinsic junctions at small twisting angles, and is suppressed by almost orders of magnitude to $45^0$ twist angle (see Fig. 2.4). At exactly $\theta = 45^0$, direct Cooper pair tunneling is forbidden due to the complete mismatch between the states with $d_{x^2-y^2}$ symmetry across the interface [11].

In Fig. 4, the values of $I_cR_N$ ($R_N$ being the normal resistivity of junction) for 24 devices investigated by Zhao et al. [12] as a function of a variable $\tilde{\theta} = \theta$ (mod $\pi/2$) at two temperatures, 12 K and 30 K, are plotted. It is observed that $I_cR_N(\tilde{\theta})$ follows $|\cos(2\tilde{\theta})|$, which is expected for somewhat incoherent Cooper pair tunneling between d-wave superconductors [11].

Martini et al. [13] report the preparation of the Josephson junction (JJ) out of the stacked $Bi_2Sr_2CuCa_2O_{8+d}$ crystals using the cryogenic dry transfer technique and encapsulating the junction with a protective insulating layer. They investigate the pairing symmetry of the Cooper pairs in the twisted BSCCO JJs by measuring the I-V characteristics and the dynamic resistance $dV/dI$ across the interface for several twisted JJs.

The current-voltage (I-V) characteristic of a representative JJ with $\theta = 43.2°$ measured at 10 K is shown in Fig. 5(a). The visible sharp SC transition across the JJs confirms significant switch-over at the interface. As the electrical current swept from a large negative value, the junction voltage $V$ first transfers to the SC state ($V = 0$) and then jumps to the resistive state at the critical current $I_c$. At low temperatures, the I-V curve of the JJ sets out a large hysteresis, associated with the high value of tunnel barrier capacitance.

Fig. 5(b) shows the angular dependence of $I_cR_N$ at two temperatures, 5 K and 30 K. It is seen that $I_cR_N$ follows a $\cos 2\theta$ dependence, which is the expected behavior of the critical current for tunneling between the d-wave superconductors in the first order approximation. The values of $I_cR_N$ in the two JJs with a tilt angle close to 45° are about two orders of magnitude lower than that in the slightly twisted JJs. The critical current is significantly reduced in these two JJs due to the nearly maximum mismatch between the wavefunctions of the two crystals, thus demonstrating the $d_{x^2-y^2}$ character of superconducting order parameters.

Lee et al. [14] report the investigation of anisotropic superconducting order parameters in twisted Bi-2212 junctions with an atomically clean vdW interface, achieved using the microcleave-and-stack technique.

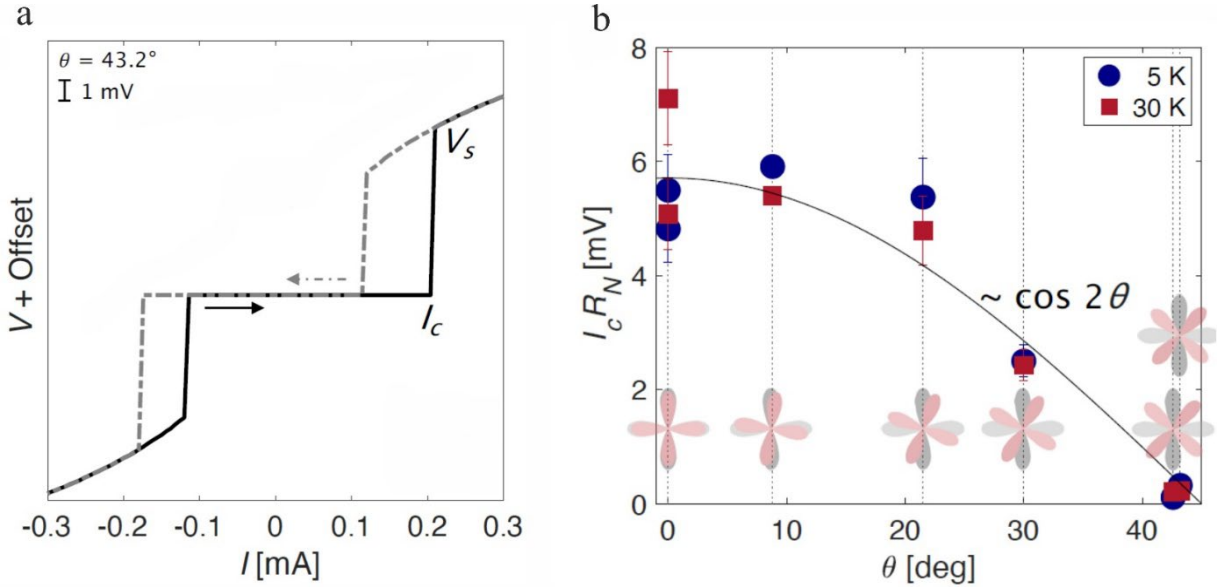

**Fig. 5.** (a) Current-voltage (I-V) characteristics of a 43.2°-twisted JJ measured by sweeping the current in both directions (arrows) at 10 K, (b) Angular dependence of $I_cR_N$ at 5 K and 30 K. The solid line follows the $\cos 2\theta$ curve. In correspondence with each marker, a schematic diagram of the d-wave wavefunctions of the crystals, twisted at the corresponding angle, is presented (after [13]).

Just like in [12] and [13] the vdW Josephson junctions with twist angles of 0° and 90° showed the maximum Josephson coupling, which was comparable to that of intrinsic Josephson (IJ) junctions in the bulk crystal. As the twist angle approaches 45°, Josephson coupling is suppressed, and eventually disappears at 45° (see Fig. 6).

copper oxide ($CuO_2$) planes where superconducting charge carriers, Cooper pairs, form. An "infinite layer" (IL) cuprate consisting essentially of $CuO_2$ planes is therefore of fundamental interest for all questions addressing the basics of high-$T_c$ superconductivity. In 1988, Siegrist et al. succeeded in synthesizing such a simple cuprate, which is known as the "parent structure" of cuprate superconductors [15]. Its $CuO_2$

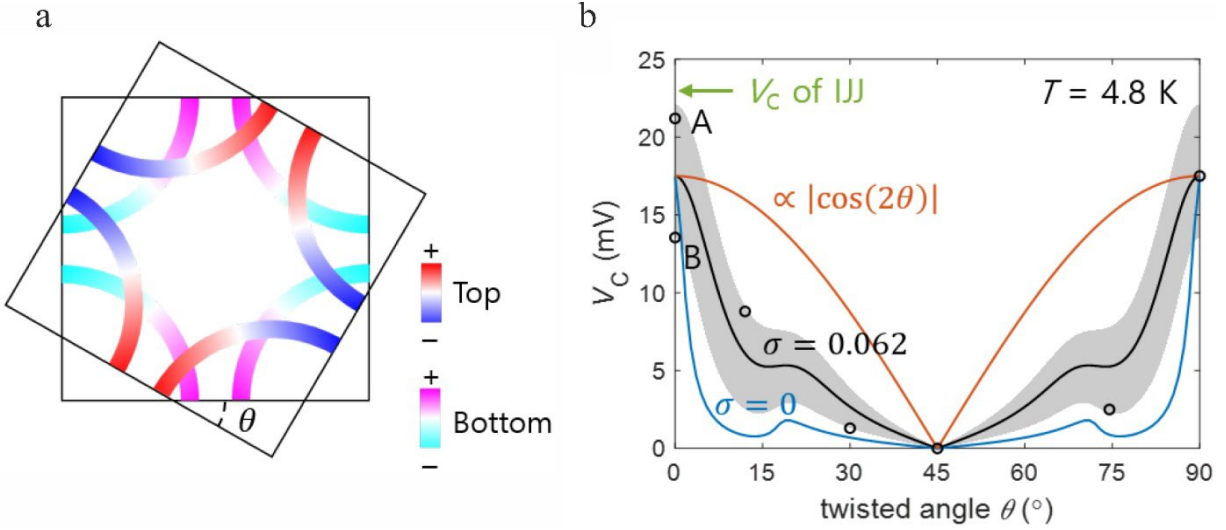

**Fig. 6.** Transport characteristics of twisted Josephson junctions with various twist angles. (a) Schematic of two Fermi surfaces from the top and bottom Bi-2212 crystals overlapping with a twist angle. The change from blue (cyan) to red (magenta) represents the sign and magnitude change of the order parameter in the top (bottom) Bi-2212 crystal. (b) Twist angle ($\theta$) dependence of the characteristic voltage ($V_c$) of twist Josephson junctions. Experimental data points are indicated as empty circles. For a full description of the figure, see the text (after [14]).

The observed twist angle dependence of the Josephson coupling was quantitatively explained by the model calculation considering the $d$-wave superconducting order, tight-binding Fermi surface of Bi-2212, and finite tunneling incoherence at the junction.

On Fig. 6 the orange line represents the $|\cos(2\theta)|$ dependence of characteristic $V_c$, theoretically expected when the circular Fermi surface is assumed. The black line is the best fitting of the experimental data using a tight-binding Fermi surface model and Gaussian tunneling model with tunneling coherence parameter $\sigma = 0.062$. The shaded region includes all data points with $\sigma$ from 0.037 to 0.071. The blue line represents the coherent limit ($\sigma = 0$) and the green line represents the $V_c$ jump of the IJ junction of the bulk Bi-2212.

planes are only separated by a single alkaline earth metal plane (A = Ca, Sr or Ba), forming an $ACuO_2$ crystal. Upon electron doping, it turned out to be superconducting with maximum $T_c$ = 43 K.

The pairing symmetry for the parent compounds up to now was essentially unknown, since a variety of experimental tests yielded conflicting results. Phase-sensitive experiments relying on Josephson junctions became available for IL cuprate thin films only very recently [16], [17].

## 3. INFINITE LAYER CUPRATE STRUCTURES

All high temperature (high-$T_c$) cuprate materials have in common that superconductivity resides in the

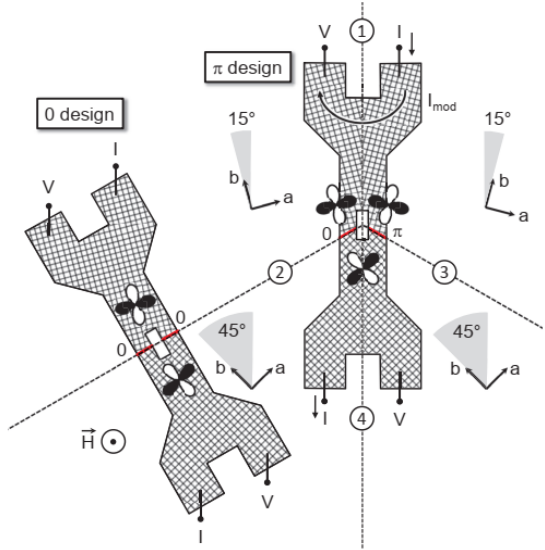

**Fig. 7.** Schematic layout of the SQUIDs. The conventional (0-design) SQUID (left) and the special (π-design) SQUID (right). For a description of the geometry of the structures, see the text. Leads for bias current *I* and voltage *V* are indicated. Magnetic fields applied perpendicular to the substrate plane (after [18]).

Tomaschko et al. [18] succeeded in depositing high quality films of the electron doped IL compound $Sr_{1-x}La_xCuO_2$ (SLCO), with $x \approx 0.15$, and in the fabrication of well-defined grain boundary Josephson junctions (GBJs) based on such SLCO films.

The authors reported on a phase sensitive study of the superconducting order parameter based on GBJ dc SQUIDs from a SLCO film grown on a tetracrystal substrate.

### 3.1. π-design SQUID

A basic device will be referred to as π-design SQUID, the SQUID ring contains one 0 junction and one π junction, if the order parameter has $d_{x^2-y^2}$-wave symmetry. The geometry both of the conventional (0-design) SQUID and of the special (π-design) SQUID, designed for identification of the $d_{x^2-y^2}$-wave pairing, is sketched in Fig. 3.1.

On Fig. 7 the $d_{x^2-y^2}$-wave order parameter is indicated by the clover-leaf structure consisting of white and black lobes, indicating the sign change of the order parameter. The grain boundary junctions (GBJ) lined with red stripes on both SQUIDs.

The 0-design SQUID includes two conventional GBJs (0 junctions) intersecting a single $30^0$ - tilt grain boundary. The π-design SQUID includes four GBJs. The misorientation angle of GB 4 is $0^0$, all other misorientation angles are $30^0$.

The measured current voltage (*IV*) characteristics of these devices could be well reproduced by the SQUID Langevin equations, extended on the nonzero junction width. A comparative analysis of data on modulation of critical current $I_c$ by magnetic field *H* for the 0-design SQUID and for the π-design SQUID was carried out.

The overall modulation of $I_c$ vs. *H* is described reasonably well by numerical calculations taking the $d_{x^2-y^2}$-wave symmetry of the order parameter into account. This is expressed qualitatively in that for the π-design SQUID $I_c$ is at a minimum near *H* = 0 - a feature which appears when one of the two Josephson junctions exhibits an additional π shift in its phase.

In summary, research on the basis of phase sensitive configuration (π - SQUID) by Tomaschko et al. [18] clearly show, that the superconducting order parameter of the electron doped infinite - layer high-$T_c$ cuprate $Sr_{1-x}La_xCuO_2$ has $d_{x^2-y^2}$-wave symmetry.

The compound $Sr_{1-x}La_xCuO_2$ has the simplest crystal structure of all high-$T_c$ cuprates ("parent structure"). Thus, it may be concluded that the $d_{x^2-y^2}$-wave symmetry is inherent to cuprate superconductivity and that it is not limited to hole doping and is not associated with the complex crystal structure of other cuprate superconductors.

### 4. ELECTRON-DOPED CUPRATES

The main results concerning the superconducting order parameter for electron-doped cuprates along with a discussion of similarities and differences in relation to the hole-doped cuprates are presented in a capital review of Armitage et al. [7]. The original generation of measurements for the order parameter symmetry in the *n*-type cuprates on polycrystals, single crystals, and thin films seemed to favor *s*-wave symmetry, but experiments on improved samples testify about the *d*-wave symmetry over most of the phase diagram with a special *nonmonotonic* functional form.

In contrast with hole-doped cuprates with simple *monotonic* $d_{x^2-y^2}$ symmetry, the interpretation of recent microwave measurements along with angle-resolved photoemission spectroscopy (ARPES), phase sensitive scanning SQUID microscope experiments and low energy electronic Raman scattering studies for the electron-doped cuprates is consistent with nonmonotonic form of the *d*-wave SC order parameter (see extensive references in [7]).

### 4.1. Nonmonotonic form of d-wave gap

Blumberg et al. [19] first proposed that a nonmonotonic *d*-wave gap function could explain the anomalous Raman response of superconducting $Nd_{2-x}Ce_xCuO_4$ (NCCO). In [19] the low energy polarized electronic Raman scattering has been investigated for single crystals of the optimally electron-doped superconductor $Nd_{2-x}Ce_xCuO_4$ ($x = 0.15$, $T_c = 22$ K). The low temperature data for different scattering channels have been found to be consistent with a *d*-wave nonmonotonic functional form of the SC gap. The gap reaches its maximum value of $4.4k_BT_c$ at Fermi surface (FS) intersections with an antiferromagnetic (AF) Brillouin zone (the "hot spots") and a smaller value of $3.3k_BT_c$ at fermionic Brillouin zone boundaries (see Fig. 8). The enhancement of the gap value in the proximity of the hot spots emphasizes the role of AF fluctuations for the superconductivity in the electron-doped cuprates.

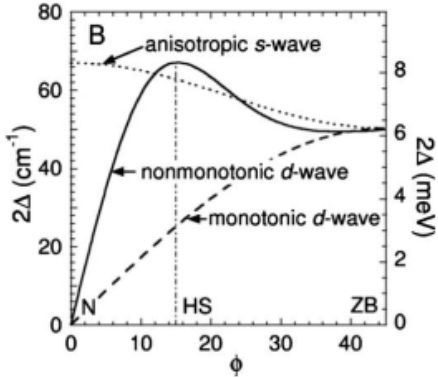

**Fig. 8.** The magnitude of SC gap as a function of the angle $\phi$ along the FS. Nonmonotonic *d*-wave gap for NCCO from analysis of Raman scattering spectra (solid line). The gap value rises rapidly from the nodal diagonal direction (N) to its maximum value at the hot spot (HS) and drops at the BZ boundary (ZB). The monotonic *d*-wave of form $\sin(2\phi)$ (dashed line) and anisotropic *s*-wave gap functions (dotted line) shown for comparison (after [19]).

Matsui *et al.* [20] performed high-resolution angle-resolved photoemission spectroscopy on electron-doped high-$T_c$ superconductor $Pr_{0.89}LaCe_{0.11}CuO_4$ (PLCCO) to study the anisotropy of the superconducting gap. Their data, shown in Fig. 9, confirmed the presence of an anisotropic gap function with zeros along the diagonal directions. The observed momentum dependence is basically consistent with the *d*-wave symmetry, but obviously deviates from the monotonic $d_{x^2-y^2}$ gap function. The maximum gap is observed not at the zone boundary, but at the hot spot where the antiferromagnetic spin fluctuation strongly couples to the electrons on the Fermi surface. The authors think that suchlike experimental results suggest the spin-mediated pairing mechanism in electron-doped high-$T_c$ superconductors.

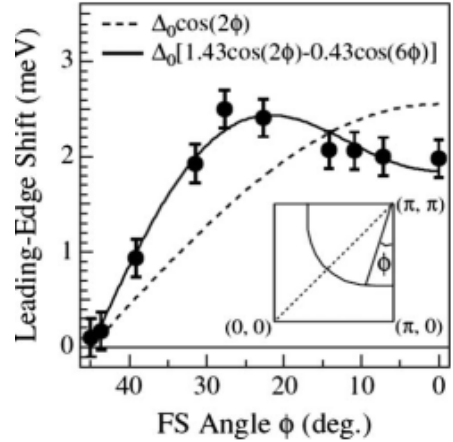

**Fig. 9.** Superconducting gap of PLCCO from ARPES. Leading-edge shift determined as a function of position (angle) on the Fermi surface showing that it fits a nonmonotonic *d*-wave symmetry (after [20]).

### 4.2. Two - band/two- gap model for NCCO

It has been first argued by Yuan et al. [21] that the experimentally observed nonmonotonic gap in electron-doped cuprates at optimal doping is the lowest quasiparticle excitation energy in the *coexisting* antiferromagnetic (AF) and superconducting (SC) state. When AF order coexists with the SC order, the resulting quasiparticle excitation can be gapped by both orders and behave to be nonmonotonic *d*-wave, even though the SC gap itself could have a typical monotonic $d_{x^2-y^2}$ symmetry (see references in [22]).

A pairing model considering AF spin fluctuation is proposed for the electron-doped cuprate superconductors by Liu and Wu [22]. It suggests that, similar to the hole-doped side, the superconducting gap function is monotonic $d_{x^2-y^2}$ wave and explains why the observed gap function has a nonmonotonic *d*-wave behavior when an AF order is taken into account.

A clue to understand the electron-doped cuprates comes from two doping-dependent FS as revealed by ARPES. Their existence is well explained in terms of the band-folding effect associated with an AF order which splits the band into upper (electron-like) and lower (hole-like) branches. In the SC state, the quasiparticles could pair each other only within the same band that leads to a two-band/two-gap model.

The situation with piecewise FS (with two types of FS parts) in superconducting state of electron - doped cuprates presented in the figures from [22] (see Fig. 10). Fig. 10a shows the FSs for a typical optimally-doped sample. The AF correlation splits the continuum FS into two pieces of sheet, in which $\alpha$ (electron-like) band is crossed by the Fermi level in the antinodal region, while $\beta$ (hole-like) band is crossed by the Fermi level in the nodal region. Fig. 10b shows an example of piecewise monotonic $d_{x^2-y^2}$ –wave $\Delta_k$, compared with a nonmonotonic one.

### 4.3. Effect of nonstoichiometric disorder on the upper critical field in NCCO

In order to establish the type of anisotropic pairing ($d$ or $s$) in electronic superconductors upon a change in the doping level we have measured temperature dependences of the resistivity in $Nd_{2-x}Ce_xCuO_{4+\delta}$ electronic superconductor, with $x = 0.14$ (underdoped region), $x = 0.15$ (optimal doping region), and $x = 0.18$ (overdoped region) and with different degrees of annealing in the oxygen-free atmosphere, in magnetic fields up to $H = 90$ kOe ($H \parallel c, J \parallel ab$) in the temperature range $T = (0.4$–$300)$ K [23]. Temperature dependences of the upper critical field $B_{c2}$ for these samples extracted from experimental data are given on Fig. 11.

The experimentally determined slope (temperature derivative near $T = T_c$) of the upper critical field, $h^* = (dB_{c2}/dT)_{T\to T_c}$, and the superconducting transition temperature $T_c/T_{c0}$ ($T_{c0}$ being the transition temperature for optimal annealing) as the functions of the degree of nonstoichiometric disorder in $Nd_{2-x}Ce_xCuO_{4+\delta}$ are presented on Fig. 12.

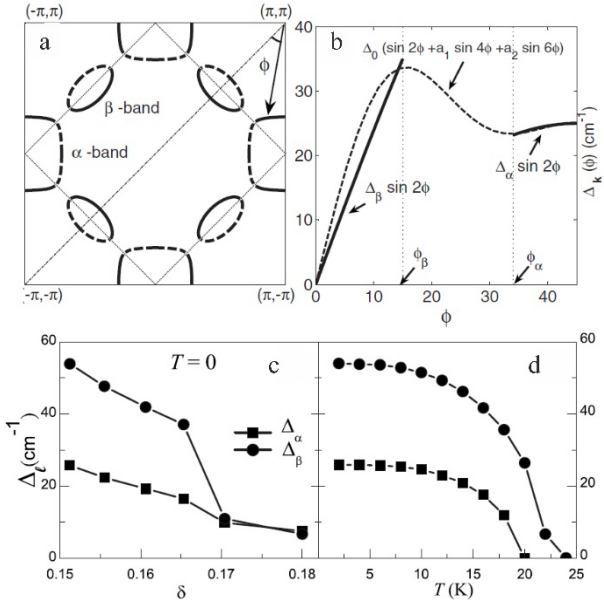

**Fig. 10.** (a) Separate $\alpha$- and $\beta$-band FSs of the electron-doped cuprates with an AF order. (b) Gap function $\Delta_k(\varphi)$ with $\varphi$ measured along the FS. Dash line: a nonmonotonic $d_{x^2-y^2}$ –wave $\Delta_k$. Solid lines: the piecewise monotonic $d_{x^2-y^2}$ -wave $\Delta_k$. The two gap amplitudes, $\Delta_\alpha$ and $\Delta_\beta$, are calculated as a function of doping (c) and temperature (d) (after [22]).

We emphasize that the model of Liu and Wu suggests the *monotonic* $d_{x^2-y^2}$ -wave SC gap function for each piece of FS (both for electron-like and hole-like parts) and explains that the observed gap function has a *nonmonotonic* $d$-wave behavior due to coexistence of superconductivity with the AF order.

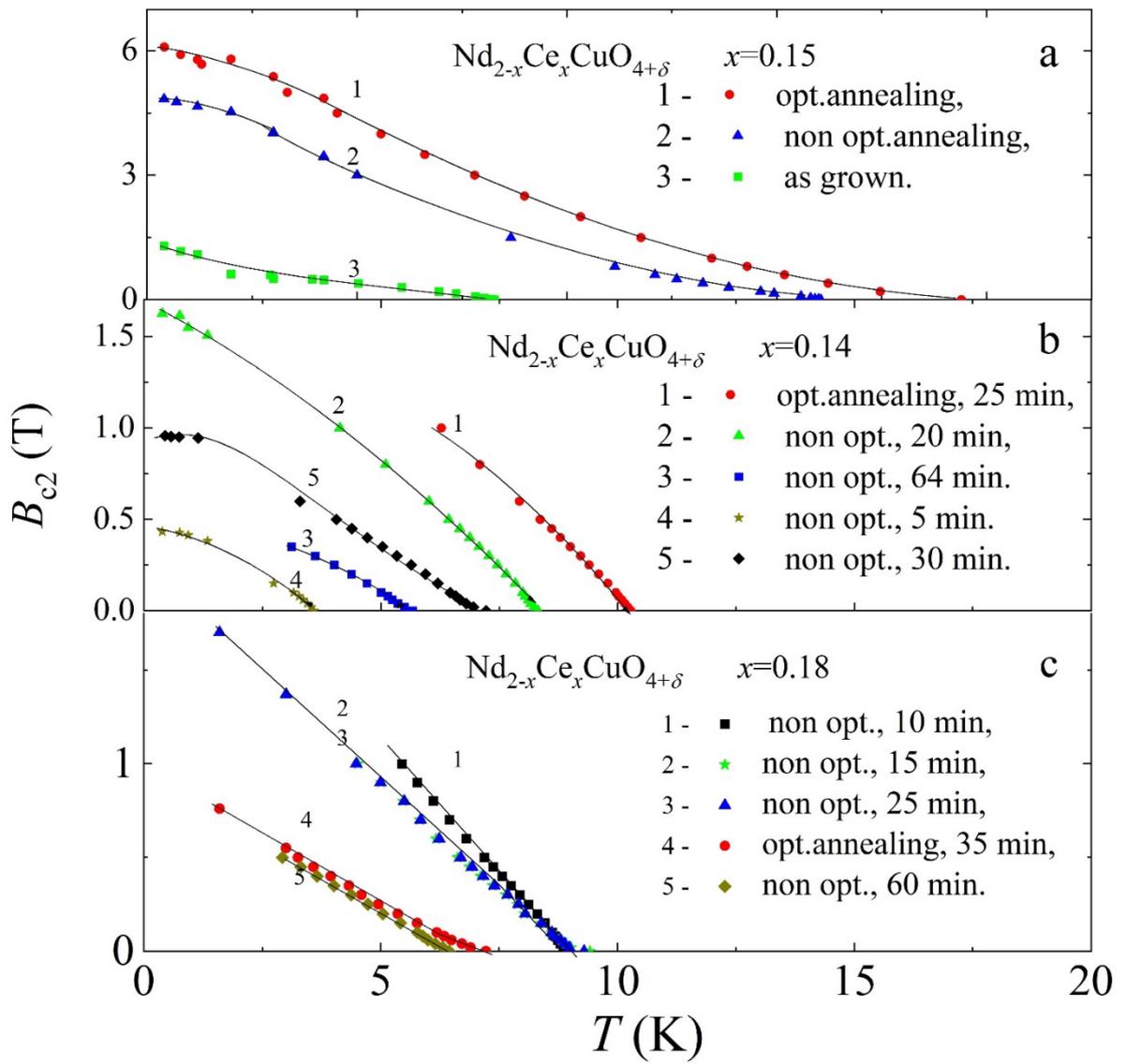

**Fig. 11.** Temperature dependencies of the upper critical field $B_{c2}$ for $Nd_{2-x}Ce_xCuO_{4+\delta}$ ((a) $x = 0.15$, (b) $x = 0.14$, (c) $x = 0.18$) electronic superconductor with different nonstoichiometric disorder (after [23]).

Effect of nonstoichiometric disorder on the upper critical field in electronic superconductors was considered theoretically in [24] where a disorder parameter $\gamma/kT_{c0}$ was introduced in the model of an impurity superconductor:

$$\gamma = \hbar/2\tau = \pi\hbar n_s/m(k_F l) \qquad (2)$$

Here $\tau$ is the electron momentum relaxation time due to scattering at normal impurities, $n_s$ is the concentration of charge carriers in the layer, $m$ is the electron effective mass, $k_F$ is the Fermi momentum and $l$ is the mean free path.

It is seen from Fig. 12(a, b) that in *optimally-doped* electronic superconductor NdCeCuO both the slope of the upper critical field and the SC transition temperature *sharply decrease* upon an increase in the degree of disorder, which is typical for systems with d-pairing [24]. A quantitative description can be achieved taking into account strong anisotropic scattering from impurities with the *d*-type symmetry (dark blue curves on figures).

In the *overdoped* NdCeCuO samples, the slope of the upper critical field *increases* with the disorder parameter, indicating that superconductivity with anisotropic *s*–type of the pairing can exist in this doping region (Fig. 12a) The observed weak dependence of SC transition temperature $T_c/T_{c0}$ on disorder parameter $\gamma/kT_{c0}$ in the overdoping region is also typical for superconductors with the *s* - type of the pairing (Fig. 12b).

In the underdoped region of the electronic superconductor, the slope of the upper critical field weakly depends on the disorder parameter (Fig. 4.5a), decreasing insignificantly when disordering becomes stronger, and the form of the dependence of $T_c/T_{c0}$ on $\gamma/kT_{c0}$ does not correspond to the theoretical calculations both for *d*-type of the pairing or for anisotropic *s*- pairing (Fig. 12b).

Thus, the analysis of the experimental temperature dependences of the resistivity of $Nd_{2-x}Ce_xCuO_{4+\delta}$ electronic superconductor has shown that in the region of optimal doping the slope of of the upper critical field decreases of the essence with an increase in the degree of disorder in the system, which, according to theoretical calculations [24], is typical for the systems with $d$- pairing.

-- Considering a mechanism induced by the AF spin fluctuation, a piecewise pairing potential is proposed for the observed nonmonotonic $d$-wave gap in optimally-doped electronic cuprate NdCeCuO [22]. The gap symmetry of electron-doped cuprate superconductors can be described properly by a two-band/two-gap model with the SC gap function being monotonic $d_{x^2-y^2}$-wave for each of the bands.

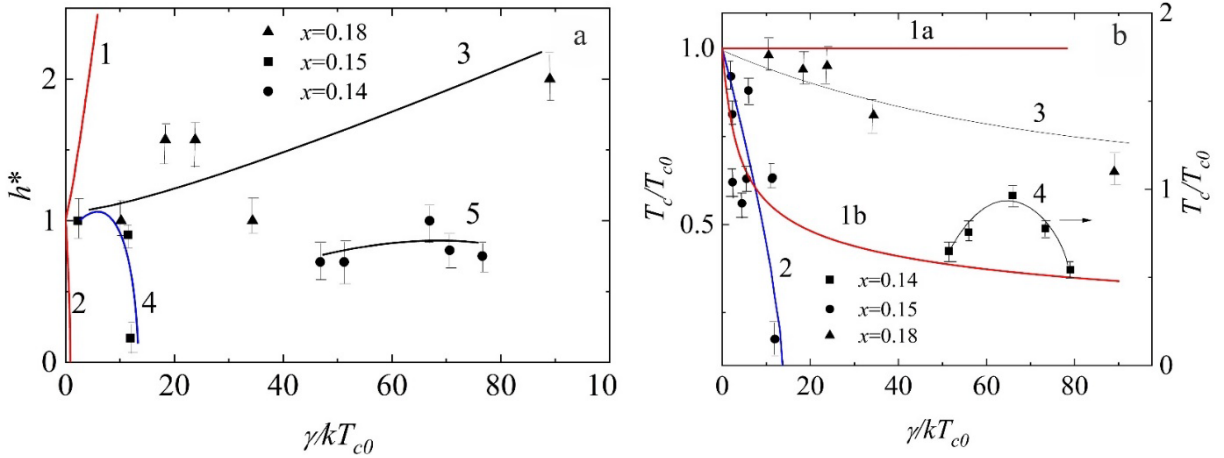

**Fig. 12.** (a) Experimental dependence of the upper critical field slope on the disorder parameter in $Nd_{2-x}Ce_xCuO_{4+\delta}$ single crystal films. The theoretical dependences for anisotropic $s$-pairing (red curve *1*), for the $d$-pairing (red curve *2*) and for the $d$ - pairing with anisotropic scattering (dark blue curve *4*) are shown for comparison. Curves *3* and *5* are guides for eyes. (b) Experimental dependence of the superconducting transition temperature on the disorder parameter in $Nd_{2-x}Ce_xCuO_{4+\delta}$ single crystal films. The theoretical dependences for isotropic $s$-pairing (red curve *1a*), for anisotropic $s$-pairing (red curve *1b*) and for the $d$-pairing with anisotropic scattering (dark blue curve *2*) are shown for comparison. Curves *3* and *4* are guides for eyes (after [23]).

## 5. CONCLUSIONS

Summing up the information presented above, the following conclusions can be drawn.

-- The study of the already established $d_{x^2-y^2}$ order parameter symmetry for hole - doped HTSC, in particular, for Bi2212 and Bi2201 systems, continues, using new high-precision techniques. Laser-based ARPES experiments with the superior resolution [8], [9] and the investigation of anisotropic superconducting order parameters in c-axis twisted Bi-2212 junctions with an atomically clean vdW interface [12], [13], [14] provide perspectives for more deep insights into the nature of energy gaps.

-- An important recent achievement is the identification of $d_{x^2-y^2}$ symmetry of superconducting order parameter for electron-doped infinite layer high-$T_c$ cuprate SrLaCuO ("parent structure") [18]. The phase sensitive configuration used was a π-design SQUID from a high quality SLCO films depositing on a tetracrystal substrate.

-- Researching an effect of nonstoichiometric disorder [23] we have found that in optimally-doped electronic cuprate NdCeCuO the slope of the upper critical field $h^* = (dB_{c2}/dT)_{T \to T_c}$ clearly decreases with an increase in the degree of disorder in the system, which is inherent in systems with $d$-wave SC pairing [24].

In the main, it seems that $d_{x^2-y^2}$-wave pairing symmetry perform as a generic type of a gap symmetry both for hole – doped and for optimally-doped electronic high-$T_c$ cuprate superconductors.